# NETWORK EXPLORATION VIA THE ADAPTIVE LASSO AND SCAD PENALTIES[1]

By Jianqing Fan, Yang Feng and Yichao Wu

*Princeton University, Princeton University and
North Carolina State University*

Graphical models are frequently used to explore networks, such as genetic networks, among a set of variables. This is usually carried out via exploring the sparsity of the precision matrix of the variables under consideration. Penalized likelihood methods are often used in such explorations. Yet, positive-definiteness constraints of precision matrices make the optimization problem challenging. We introduce nonconcave penalties and the adaptive LASSO penalty to attenuate the bias problem in the network estimation. Through the local linear approximation to the nonconcave penalty functions, the problem of precision matrix estimation is recast as a sequence of penalized likelihood problems with a weighted $L_1$ penalty and solved using the efficient algorithm of Friedman et al. [*Biostatistics* **9** (2008) 432–441]. Our estimation schemes are applied to two real datasets. Simulation experiments and asymptotic theory are used to justify our proposed methods.

**1. Introduction.** Network modeling is often explored via estimating the sparse precision matrix, the inverse covariance matrix, in which off-diagonal elements represent the conditional covariance between the corresponding variables. The sparsity is often studied via penalized likelihood, with an appropriately chosen penalty function. The results are usually summarized graphically by linking conditionally dependent variables. This provides an understanding of how variables, such as the coexpression of genes, are related to each other. A challenge in network modeling is to optimize the penalized likelihood, subject to the positive-definiteness constraint of the precision matrix. Further challenges arise in reducing the biases induced by the penalized likelihood method.

---

Received November 2007; revised October 2008.

[1]Supported in part by NIH Grant R01-GM072611 and NSF Grants DMS-07-04337 and DMS-07-14554.

Let $\mathbf{X} = (X_1, X_2, \ldots, X_p)^T$ be a $p$-dimension random vector having a multivariate normal distribution with mean vector $\boldsymbol{\mu}$ and covariance matrix $\boldsymbol{\Sigma}$. The research on large covariance matrix estimation has surged recently due to high-dimensional data, generated by modern technologies such as microarray, fMRI and so on. In many applications like gene classifications and optimal portfolio allocations it is the precision matrix, denoted by $\boldsymbol{\Omega} \equiv \boldsymbol{\Sigma}^{-1}$, that is needed and plays an important role. It has a nice interpretation in the Gaussian graphical model, as the $(i,j)$-element of $\boldsymbol{\Omega}$ is exactly the partial correlation between the $i$th and $j$th components of $\mathbf{X}$. In the Gaussian concentration graphical model with undirected graph $(V, E)$, vertices $V$ correspond to components of the vector $\mathbf{X}$ and edges $E = \{e_{ij}, 1 \leq i, j \leq p\}$ indicate the conditional dependence among different components of $\mathbf{X}$. The edge $e_{ij}$ between $X_i$ and $X_j$ exists if and only if $\omega_{ij} \neq 0$, where $\omega_{ij}$ is the $(i,j)$-element of $\boldsymbol{\Omega}$. Hence, of particular interest is to identify null entries in the precision matrix.

There is significant literature on model selection and parameter estimation in the Gaussian concentration graphical model. The seminal paper by Dempster (1972) discussed the idea of simplifying the covariance structure by setting some elements of the precision matrix to zero. Initially the methods of precision matrix estimation were based on two steps: (1) identify the "correct" model; (2) estimate the parameters for the identified model. One standard approach for identifying the model is the greedy stepwise forward-selection (or backward-selection), which is achieved through hypothesis testing; see Edwards (2000) for an extensive introduction. Drton and Perlman (2004) noted that it is not clear whether the stepwise method is valid as a simultaneous testing procedure because its overall error rate is not controlled. To improve this stepwise method, Drton and Perlman (2004) proposed a conservative simultaneous confidence interval to select model in a single step. Using the least absolute shrinkage and selection operator (LASSO) [Tibshirani (1996)], Meinshausen and Bühlmann (2006) proposed to perform a neighborhood selection at each node in the graph. This neighborhood selection is computationally very fast and suitable for large-size problems.

The instability of the aforementioned two-step procedures has been recognized by Breiman (1996). Fan and Li (2001) proposed the penalized likelihood, which can achieve model selection and parameter estimation simultaneously. This penalized likelihood was later studied by d'Aspremont, Banerjee and Ghaoui (2008), Yuan and Lin (2007), Levina, Zhu and Rothman (2008), Rothman et al. (2008) and Friedman, Hastie and Tibshirani (2008) in the context of precision matrix estimation. Yuan and Lin (2007) solved the corresponding optimization problem using the MAXDET algorithm [Vandenberghe, Boyd and Wu (1998)] and focused on statistical properties of the estimates. d'Aspremont, Banerjee and Ghaoui (2008) proposed two efficient first-order numerical algorithms with low memory requirement using semidefinite programming algorithms, which obey the positive-definiteness constraint of the precision matrix. Rothman et al. (2008) and Lam and Fan (2008) showed that the Frobenius norm between the inverse correlation matrix and its $L_1$ penalized likelihood estimator is $O_p(\sqrt{S \log p/n})$, where $S$ is the number of the nonzero elements of the inverse of the correlation matrix. Consequently, the sparse inverse correlation matrix is highly estimable and the dimensionality only costs an order of $\log p$, a remarkable improvement on the general result of Fan and Peng (2004). Using a coordinate descent procedure, Friedman, Hastie and Tibshirani (2008) proposed the graphical lasso algorithm to estimate the sparse inverse covariance matrix using the LASSO penalty. The graphical lasso algorithm is remarkably fast.

The $L_1$ penalty is convex and leads to a desirable convex optimization problem when the log-likelihood function is convex. Recent innovation of the LARS algorithm [Efron et al. (2004)] enables computation of the whole solution path of the $L_1$ penalized regression within $O(n^2 p)$ operations. This is a remarkable achievement. However, such an algorithm does not apply to the estimation of the precision matrix, whose parameters are subject to a positive-definiteness constraint of the matrix.

It has been shown that the LASSO penalty produces biases even in the simple regression setting [Fan and Li (2001)] due to the linear increase of the penalty on regression coefficients. To remedy this bias issue, two new penalties were proposed recently: one is the nonconcave penalty, such as the Smoothly Clipped Absolute Deviation (SCAD) penalty [Fan and Li (2001)], and the other is the adaptive LASSO penalty due to Zou (2006). In this work we will study precision matrix estimation using these two penalty functions. Lam and Fan (2008) studied theoretical properties of sparse precision matrices estimation via a general penalty function satisfying the properties in Fan and Li (2001). The bias presented in the LASSO penalty is also demonstrated for sparse precision matrix estimation in Lam and Fan (2008). Through the local linear approximation [Zou and Li (2008)] to the nonconcave penalty function, the nonconcave penalized likelihood can be recast as a sequence of weighted $L_1$ penalized likelihood problems. The weighting scheme is governed by the derivative of the penalty function, which depends on the magnitude of the current estimated coefficient: the larger magnitude the smaller weight. Therefore, the optimization of the penalized likelihood with a nonconcave penalty subject to the positive-definiteness constraint of $\Omega$ can be elegantly solved by the efficient algorithm of Friedman, Hastie and Tibshirani (2008). In this way, we simultaneously solve the bias issue and reduce the computational burden.

Other recent work on Gaussian concentration graphical models includes the following: Li and Gui (2006), who introduced a threshold gradient descent (TGD) regularization procedure for the sparse precision matrix estimation; Schäfer and Strimmer (2005), who estimated the correlation matrix via regularization with bootstrap variance reduction and used false discovery rate multiple testing to select network based on the estimated correlation matrix; Bayesian approaches considered in Wong, Carter and Kohn (2003) and Dobra et al. (2004); Huang et al. (2006), who reparameterized the covariance matrix through the modified Cholesky decomposition of its inverse and transferred covariance matrix estimation to the task of model selection and estimation for a sequence of regression models, among others.

The rest of the paper is organized as follows. Section 2 describes the algorithm for precision matrix estimation and three types of penalties in detail. In Section 3 our methods are applied to two real datasets: telephone call center data [Shen and Huang (2005)] and pCR development of breast cancer [Hess et al. (2006)]. Section 4 uses Monte Carlo simulation to compare the performance of the three kinds of penalty functions under consideration. Theoretical properties of the SCAD and adaptive LASSO penalized approach are used to justify our methods in Section 5. The Appendix collects all the technical proofs.

**2. Methods.** Suppose $\mathbf{x}_1, \mathbf{x}_2, \ldots, \mathbf{x}_n$ are from a Gaussian distribution with unknown mean vector $\boldsymbol{\mu}_0$ and covariance matrix $\boldsymbol{\Sigma}_0$, denoted as $N(\boldsymbol{\mu}_0, \boldsymbol{\Sigma}_0)$, where $\mathbf{x}_i = (x_{i1}, x_{i2}, \ldots, x_{ip})^T$. Denote the sample covariance matrix by $\hat{\boldsymbol{\Sigma}}$, whose $(j,k)$-element $\hat{\sigma}_{jk}$ is given by $\sum_{i=1}^{n}(x_{ij} - \bar{x}_j)(x_{ik} - \bar{x}_k)/n$, where $\bar{x}_j = \sum_{i=1}^{n} x_{ij}/n$ is the sample mean of the $j$th component. Note that we use $n$ instead of $n - p$ in the definition of the sample covariance matrix so that the log-likelihood function of the precision matrix can be written in a compact format as in (2.1).

2.1. *Penalized likelihood estimation.* The precision matrix $\boldsymbol{\Omega} = \boldsymbol{\Sigma}^{-1}$ is estimated by maximizing twice the log-likelihood function, which is given by

$$(2.1) \qquad 2l(\boldsymbol{\Omega}) = \log \det \boldsymbol{\Omega} - \langle \hat{\boldsymbol{\Sigma}}, \boldsymbol{\Omega} \rangle + \text{Constant},$$

where $\langle \hat{\boldsymbol{\Sigma}}, \boldsymbol{\Omega} \rangle = \text{tr}(\hat{\boldsymbol{\Sigma}} \boldsymbol{\Omega})$ denotes the trace of the product matrix $\hat{\boldsymbol{\Sigma}} \boldsymbol{\Omega}$. When $n > p$, the global maximizer of $l(\boldsymbol{\Omega})$ is given by $\hat{\boldsymbol{\Omega}} = \hat{\boldsymbol{\Sigma}}^{-1}$.

Denote the generic penalty function on each element by $p(\cdot)$. Under the penalized likelihood framework, the estimate of the sparse precision matrix is the solution to the following optimization problem:

$$(2.2) \qquad \max_{\boldsymbol{\Omega} \in S_p} \log \det \boldsymbol{\Omega} - \langle \hat{\boldsymbol{\Sigma}}, \boldsymbol{\Omega} \rangle - \sum_{i=1}^{p} \sum_{j=1}^{p} p_{\lambda_{ij}}(|\omega_{ij}|),$$

where $\omega_{ij}$ is the $(i,j)$-element of matrix $\boldsymbol{\Omega}$ and $\lambda_{ij}$ is the corresponding tuning parameter.

The LASSO penalty proposed by Tibshirani (1996) achieves sparsity in the regression setting. Essentially, the LASSO penalty uses the $L_1$ penalty function: $p_\lambda(|x|) = \lambda|x|$. Friedman, Hastie and Tibshirani (2008) applied the LASSO penalty to (2.2) and proposed the graphical lasso algorithm by using a coordinate descent procedure, which is remarkably fast. Moreover, this algorithm allows a "warm" start, from which we can use the estimate for one value of the tuning parameter as the starting point for the next value.

Numerical examples show that the LASSO penalty can produce a sparse estimate of the precision matrix. However, the LASSO penalty increases linearly in the magnitude of its argument. As a result, it produces substantial biases in the estimates for large regression coefficients. To address this issue, Fan and Li (2001) proposed a unified approach via nonconcave penalties. They gave necessary conditions for the penalty function to produce sparse solutions, to ensure consistency of model selection, and to result in unbiased estimates for large coefficients. All three of these desirable properties are simultaneously achieved by the SCAD penalty, proposed by Fan (1997). Mathematically, the SCAD penalty is symmetric and a quadratic spline on $[0, \infty)$, whose first order derivative is given by

$$(2.3) \qquad \text{SCAD}'_{\lambda,a}(x) = \lambda \left\{ I(|x| \le \lambda) + \frac{(a\lambda - |x|)_+}{(a-1)\lambda} I(|x| > \lambda) \right\}$$

for $x \ge 0$, where $\lambda > 0$ and $a > 2$ are two tuning parameters. When $a = \infty$, (2.3) corresponds to the $L_1$ penalty. Based on an argument of minimizing the Bayes risk, Fan and Li (2001) recommended the choice $a = 3.7$, which will be used in all of our numerical examples. Using the SCAD penalty, we are seeking to solve the following optimization problem:

$$(2.4) \qquad \max_{\boldsymbol{\Omega} \in S_p} \log \det \boldsymbol{\Omega} - \langle \hat{\boldsymbol{\Sigma}}, \boldsymbol{\Omega} \rangle - \sum_{i=1}^{p} \sum_{j=1}^{p} \text{SCAD}_{\lambda,a}(|\omega_{ij}|),$$

where we set $\lambda_{ij} = \lambda$ for convenience.

Zou (2006) proposed another method to achieve the aforementioned three desirable properties simultaneously. It is called the adaptive LASSO penalty, and requires a weight for each component. The adaptive LASSO penalty is essentially a weighed version of the LASSO penalty with some properly chosen weights. For our setting, we define the adaptive weights to be $w_{ij} = 1/|\tilde{\omega}_{ij}|^\gamma$ for some $\gamma > 0$ and any consistent estimate $\tilde{\boldsymbol{\Omega}} = (\tilde{\omega}_{ij})_{1 \le i, j \le p}$. Putting the adaptive LASSO penalty into (2.2), we get

$$(2.5) \qquad \max_{\boldsymbol{\Omega} \in S_p} \log \det \boldsymbol{\Omega} - \langle \hat{\boldsymbol{\Sigma}}, \boldsymbol{\Omega} \rangle - \lambda \sum_{i=1}^{p} \sum_{j=1}^{p} w_{ij} |\omega_{ij}|.$$

This method was proposed by Zou (2006) in the regression setting. According to our numerical experience, estimation results do not differ much for different $\gamma$. So, for simplicity, we fix $\gamma = 0.5$ in all our numerical analysis. Furthermore, the initial estimate $\hat{\boldsymbol{\Omega}}$ can be chosen as the inverse sample covariance matrix for the case $p < n$ or the precision matrix estimate derived from the LASSO penalty for the case $p \geq n$. Note that the inverse sample covariance matrix when $p < n$ may not be consistent if we allow $p$ to grow with $n$. This requirement of a consistent initial estimate is a drawback of the adaptive LASSO. In the next subsection we elucidate the connection of the nonconcave penalty to the adaptive LASSO penalty.

2.2. *Iterative reweighted penalized likelihood.* To reduce the biases for estimating nonzero components, Fan and Li (2001) pointed out a necessary condition that the penalty function $p_\lambda(\cdot)$ should be nondecreasing over $[0, \infty)$ while leveling off near the tail. Hence, the penalty function needs to be concave on $[0, \infty)$. At the time, in the absence of the innovative LARS algorithm [Efron et al. (2004)], they proposed the LQA algorithm, which conducts the optimization iteratively and in each step approximates the SCAD penalty via a quadratic function. Hunter and Li (2005) studied the LQA in a more general framework in terms of the MM (minorize–maximize) algorithm and showed its nice asymptotic properties. The SPICE of Rothman et al. (2008) is also based on the LQA algorithm. For both the LQA and MM algorithms, Friedman, Hastie and Tibshirani (2008)'s graphical lasso algorithm cannot directly be applied because the penalty is locally approximated by a quadratic function.

In this work, to take advantage of the graphical lasso algorithm of Friedman, Hastie and Tibshirani (2008), we resort to the local linear approximation (LLA), proposed in Zou and Li (2008), which is an improvement of the LQA in Fan and Li (2001). In each step, the LLA algorithm locally approximates the SCAD penalty by a symmetric linear function. For any $\omega_0$, by the Taylor expansion, we approximate $p_\lambda(|\omega|)$ in a neighborhood of $|\omega_0|$ as follows:

$$p_\lambda(|\omega|) \approx p_\lambda(|\omega_0|) + p_\lambda'(|\omega_0|)(|\omega| - |\omega_0|),$$

where $p_\lambda'(\omega) = \frac{\partial}{\partial \omega} p_\lambda(\omega)$, which is nonnegative for $\omega \in [0, \infty)$ due to the monotonicity of $p_\lambda(\cdot)$ over $[0, \infty)$. Denote the $k$-step solution by $\hat{\boldsymbol{\Omega}}^{(k)}$. Consequently, at step $k$, we are optimizing, up to a constant,

$$(2.6) \qquad \max_{\boldsymbol{\Omega} \in S_p} \log \det \boldsymbol{\Omega} - \langle \hat{\boldsymbol{\Sigma}}, \boldsymbol{\Omega} \rangle - \sum_{i=1}^{p} \sum_{j=1}^{p} w_{ij} |\omega_{ij}|,$$

where $w_{ij} = p_\lambda'(|\hat{\omega}_{ij}^{(k)}|)$ and $\hat{\omega}_{ij}^{(k)}$ is the $(i,j)$-element of $\hat{\boldsymbol{\Omega}}^{(k)}$. The optimization problem (2.6) can be easily solved by the graphical lasso algorithm proposed by Friedman, Hastie and Tibshirani (2008).

At each step, (2.6) is equivalent to a weighted version of the $L_1$-penalized likelihood, leading to a sparse solution. The weighting scheme is governed by the derivative of the penalty function and the magnitude of the current estimate: the larger magnitude the smaller weight. In Theorem 5.1, we show that the penalized likelihood objective function is increasing through each iteration in the LLA algorithm. Due to the sparsity in each iteration, Zou and Li (2008) studied the one-step LLA algorithm and showed that, asymptotically, the one-step algorithm performs as well as the fully iterative LLA algorithm as long as the initial solution is good enough. As a result, we simply use the one-step LLA algorithm in this work. In our implementation, the initial value is taken as either the inverse sample covariance matrix or the LASSO estimate of the precision matrix. The latter is equivalent to using (2.6) twice starting with the primitive initial value $\hat{\boldsymbol{\Omega}}^{(0)} = 0$, resulting in the LASSO estimate $\hat{\boldsymbol{\Omega}}^{(1)}$ in the first step as $\mathrm{SCAD}'_{\lambda,a}(0) = \lambda$. This also demonstrates the flexibility of the SCAD penalty: an element being estimated as zero can escape from zero in the next iteration, whereas the adaptive LASSO absorbs zeros in each application (the estimate is always sparser than the initial value).

2.3. *Tuning parameter selection.* As in every regularization problem, the tuning parameter $\lambda$ controls the model complexity and has to be tuned for each penalty function. In this work we use the popular $K$-fold cross-validation method to do the tuning parameter selection. First divide all the samples in the training dataset into $K$ disjoint subgroups, also known as folds, and denote the index of subjects in the $k$th fold by $T_k$ for $k = 1, 2, \ldots, K$. The $K$-fold cross-validation score is defined as

$$CV(\lambda) = \sum_{k=1}^{K} \left( n_k \log |\hat{\boldsymbol{\Omega}}_{-k}(\lambda)| - \sum_{i \in T_k} (\mathbf{x}^{(i)})^T \hat{\boldsymbol{\Omega}}_{-k}(\lambda) \mathbf{x}^{(i)} \right),$$

where $n_k$ is the size of the $k$th fold $T_k$ and $\hat{\boldsymbol{\Omega}}_{-k}(\lambda)$ denotes the estimate of the precision matrix based on the sample $(\bigcup_{k=1}^{K} T_k) \backslash T_k$ with $\lambda$ as the tuning parameter. Then, we choose $\lambda^* = \arg\max_\lambda CV(\lambda)$ as the best tuning parameter, which is used to obtain the final estimate of the precision matrix based on the whole training set $\bigcup_{k=1}^{K} T_k$. Here the maximization of $CV(\lambda)$ with respect to $\lambda$ is achieved via a grid search.

**3. Application to real data.** In this section we apply our estimation scheme to two real datasets and compare the performance of three different penalty functions: the LASSO, adaptive LASSO and SCAD.

3.1. *Telephone call center data.* In this example our method is applied to forecast the call arrival pattern of a telephone call center. The data come from one call center in a major U.S. northeastern financial organization, containing the information about the arrival time of every call at the service queue. Phone calls are recorded from 7:00AM until midnight for each day in 2002, except 6 days when the data-collecting equipment was out of order. More details about this data can be found in Shen and Huang (2005).

We take the same data preprocessing as in Huang et al. (2006): (1) divide the 17-hour period into 102 10-minute intervals; (2) count the number of calls arriving at the service queue during each interval; (3) focus on weekdays only; (4) use the singular value decomposition to screen out outliers that include holidays and days when the recording equipment was faulty. Finally, we have observations for 239 days. Denote the data for day $i$ by $\mathbf{N}_i = (N_{i1}, \ldots, N_{i,102})'$, for $i = 1, \ldots, 239$, where $N_{it}$ is the number of calls arriving at the call center for the $t$th 10-minute interval on day $i$. Define $y_{it} = \sqrt{N_{it} + 1/4}$ using the variance stabilization transform for $i = 1, \ldots, 239$ and $t = 1, \ldots, 102$. We apply the penalized likelihood estimation method with three different penalty functions: the LASSO, adaptive LASSO and SCAD, to estimate the $102 \times 102$ precision matrix. As in Huang et al. (2006), we use the estimated precision matrix to forecast the number of arrivals later in the day using arrival patterns at earlier times of the day. Denote $\mathbf{y}_i = (y_{i1}, \ldots, y_{i,102})'$. Then form the partition $\mathbf{y}_i = (\mathbf{y}_i^{(1)'}, \mathbf{y}_i^{(2)'})'$, where $\mathbf{y}_i^{(1)}$ and $\mathbf{y}_i^{(2)}$ represent the arrival patterns in the early and the later time of day $i$. Here we can take $\mathbf{y}_i^{(1)} = (y_{i1}, \ldots, y_{i,51})'$ and $\mathbf{y}_i^{(2)} = (y_{i,52}, \ldots, y_{i,102})'$. The corresponding partition of the mean and covariance matrix is

$$\boldsymbol{\mu} = \begin{pmatrix} \boldsymbol{\mu}_1 \\ \boldsymbol{\mu}_2 \end{pmatrix}, \qquad \boldsymbol{\Sigma} = \begin{pmatrix} \boldsymbol{\Sigma}_{11}, \boldsymbol{\Sigma}_{12} \\ \boldsymbol{\Sigma}_{21}, \boldsymbol{\Sigma}_{22} \end{pmatrix}.$$

With the multivariate normality assumption, the best mean squared error forecast of $\mathbf{y}_i^{(2)}$ using $\mathbf{y}_i^{(1)}$ is

$$E(\mathbf{y}_i^{(2)} | \mathbf{y}_i^{(1)}) = \boldsymbol{\mu}_2 + \boldsymbol{\Sigma}_{21} \boldsymbol{\Sigma}_{11}^{-1} (\mathbf{y}_i^{(1)} - \boldsymbol{\mu}_1),$$

which is also the best linear predictor for non-Gaussian data.

To evaluate the forecasting performance, we split the 239 days into training and testing days. The data from the first 205 days, corresponding from January to October, is used as the training dataset to estimate the mean vector $\boldsymbol{\mu}$ and the precision matrix $\boldsymbol{\Omega} = \boldsymbol{\Sigma}^{-1}$. The remaining 34 days are used for testing. We define the average absolute forecast error (AAFE) by

$$\text{AAFE}_t = \frac{1}{34} \sum_{i=206}^{239} |\hat{y}_{it} - y_{it}|,$$

TABLE 1
*Average result of call center prediction*

|                                         | **Sample** | **LASSO** | **Adaptive LASSO** | **SCAD** |
|-----------------------------------------|------------|-----------|--------------------|----------|
| Average AAFE                            | 1.46       | 1.39      | 1.34               | 1.31     |
| Nonzero elements in $\hat{\mathbf{\Sigma}}_{11}^{-1}$ | 10,394     | 2788      | 1417               | 684      |

where $y_{it}$ and $\hat{y}_{it}$ are the observed and the predicted values, respectively. In Figure 1 we compare the AAFE performance using the sample covariance matrix and the penalized estimates with the LASSO, adaptive LASSO and SCAD penalties. In Table 1 we give the average AAFE of the 34 days we set aside for testing and also the number of the nonzero elements in the precision matrix estimate of the four methods. Here and in all following numerical studies, we let the element $\omega_{ij}$ of the precision matrix be zero if $|\omega_{ij}| < 10^{-3}$, because the default threshold for convergence in the graphical lasso algorithm is $10^{-4}$. We have tried several other thresholding levels, such as $10^{-2}$ and $10^{-4}$, and obtained similar conclusions in both real data analysis and simulations.

Figure 1 and Table 1 show clearly that the forecasts based on the penalized estimates are better than that based on the sample covariance matrix. Among the three penalized estimates, the estimate associated with the SCAD penalty performs the best, followed by the adaptive LASSO, and finally the LASSO forecast. Moreover, we can see that the sample precision matrix is a nonsparse precision matrix and leads to a much more complex network than the penalized ones. Comparing to the LASSO, the adaptive LASSO leads to a simpler network and the SCAD provides an even simpler network, resulting in the smallest forecasting errors. The reason is that the SCAD penalty results in the least biased estimate among three penalized schemes. This allows the data to choose a larger penalty parameter $\lambda$ for the SCAD penalty and set more spurious zeros to zero. This phenomenon will also be observed and demonstrated in the simulation studies.

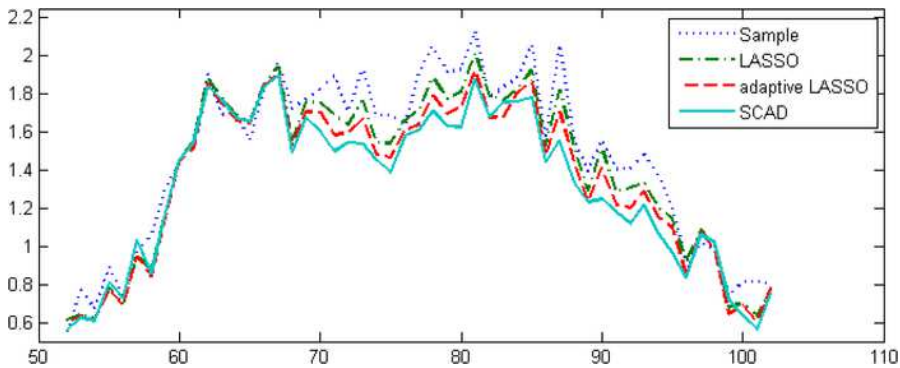

FIG. 1. *Average absolute forecast error $AAAE_t$ against $t = 52, \ldots, 102$ using the sample estimate and using three penalties: LASSO, adaptive LASSO and SCAD.*

3.2. *Breast cancer data.* As a second example, we focus on selecting gene expression profiling as a potential tool to predict the breast cancer patients who may achieve pathologic Complete Response (pCR), which is defined as no evidence of viable, invasive tumor cells left in the surgical specimen. As in Kuerer et al. ([1999](#)), pCR after neoadjuvant chemotherapy has been described as a strong indicator of survival, justifying its use as a surrogate marker of chemosensitivity. Consequently, considerable interest has been developed in finding methods to predict which patients will have a pCR to preoperative therapy. In this study we use the normalized gene expression data of 130 patients with stages I–III breast cancers analyzed by Hess et al. ([2006](#)). Among the 130 patients, 33 of them are from class 1 (achieved pCR), while the other 97 belong to class 2 (did not achieve pCR).

To evaluate the performance of the penalized precision matrix estimation using three different penalties, we randomly divide the data into training and testing sets of sizes 109 and 21, respectively, and repeat the whole process 100 times. To maintain similar class proportion for the training and testing datasets, we use a stratified sampling: each time we randomly select 5 subjects from class 1 and 16 subjects from class 2 (both are roughly 1/6 of their corresponding total class subjects) and these 21 subjects make up the testing set; the remaining will be used as the training set. From each training data, we first perform a two-sample $t$-test between the two groups and select the most significant 110 genes that have the smallest $p$-values. In this case, the dimensionality $p = 110$ is slightly larger than the sample size $n = 109$ for training datasets in our classification study. Due to the noise accumulation demonstrated in Fan and Fan ([2008](#)), $p = 110$ may be larger than needed for optimal classification, but allows us to examine the performance when $p > n$. Second, we perform a gene-wise standardization by dividing the data with the corresponding standard deviation, estimated from the training dataset. Finally, we estimate the precision matrix and consider the linear discriminant analysis (LDA). LDA assumes that the normalized gene expression data in class-$k$ is normally distributed as $N(\boldsymbol{\mu}_k, \boldsymbol{\Sigma})$ with the same covariance matrix, where $k = 1, 2$. The linear discriminant scores are as follows:

$$\delta_k(\mathbf{x}) = \mathbf{x}^T \hat{\boldsymbol{\Sigma}}^{-1} \hat{\boldsymbol{\mu}}_k - \tfrac{1}{2} \hat{\boldsymbol{\mu}}_k^T \hat{\boldsymbol{\Sigma}}^{-1} \hat{\boldsymbol{\mu}}_k + \log \hat{\pi}_k,$$

where $\hat{\pi}_k = n_k/n$ is the proportion of the number of observations in the training data belonging to the class $k$, and the classification rule is given by $\arg\max_k \delta_k(\mathbf{x})$. Details for LDA can be found in Mardia, Kent and Bibby

(1979). Based on each training dataset, we can estimate the with-in class mean vectors by

$$\hat{\boldsymbol{\mu}}_k = \frac{1}{n_k} \sum_{i \in \text{class-}k} \mathbf{x}_i \qquad \text{for } k = 1, 2$$

and precision matrix $\boldsymbol{\Sigma}^{-1}$ using the penalized loglikelihood method with three different penalty functions: the LASSO, adaptive LASSO and SCAD. Tuning parameters in different methods are chosen via six-fold cross-validation based on the training data. Note that the sample size $n$ is smaller than the dimensionality $p$ in this case. As a result, the sample covariance matrix is degenerate and cannot be used in the LDA.

To compare the prediction performance, we used specificity, sensitivity and also Matthews Correlation Coefficient(MCC). They are defined as follows:

$$\text{Specificity} = \frac{\text{TN}}{\text{TN} + \text{FP}}, \qquad \text{Sensitivity} = \frac{\text{TP}}{\text{TP} + \text{FN}},$$

$$\text{MCC} = \frac{\text{TP} \times \text{TN} - \text{FP} \times \text{FN}}{\sqrt{(\text{TP} + \text{FP})(\text{TP} + \text{FN})(\text{TN} + \text{FP})(\text{TN} + \text{FN})}},$$

where TP, TN, FP and FN are the numbers of true positives, true negatives, false positives and false negatives, respectively. MCC is widely used in machine learning as a measure of the quality of binary classifiers. It takes true and false, positives and negatives, into account and is generally regarded as a balanced measure, which can be used even if the classes are of very different sizes. The larger the MCC is, the better the classification is. More details can be found in Bladi et al. (2000). Means and standard deviations (in parentheses) of the specificity, sensitivity, MCC and the number of nonzero elements in $\hat{\boldsymbol{\Omega}}$ over 100 repetitions are reported in Table 2. To visually interpret the gene network derived by our penalized likelihood methods, we applied our whole estimation scheme to all the 130 datasets: (1) use a two sample $t$-test to select 110 genes; (2) use the penalized likelihood estimation scheme to derive the precision matrix estimates. Next we try to show the corresponding gene networks derived by using three different penalties. To gain a better view, we only plot the gene networks of the 60 genes with the smallest p-values among the 110 genes in Figure 2.

From the table, we can see that the adaptive LASSO and SCAD improve over the LASSO in terms of the specificity and MCC, while all three penalties give similar sensitivity. Furthermore, when we look at the number of nonzero elements of the precision matrix estimates, using three different penalties, we can see again that, by using the adaptive LASSO and SCAD penalties, we can get much simpler models which are often more desirable. From Figure 2, it is clear that, compared with the network derived using the LASSO penalty, the ones derived using the adaptive LASSO and SCAD penalties both show some small clusters, indicating block diagonal precision matrices. This interesting phenomenon is worth further study.

TABLE 2
*Result of pCR classification over 100 repetitions*

|                | Specificity    | Sensitivity    | MCC            | Nonzero elements in $\hat{\Omega}$ |
|----------------|----------------|----------------|----------------|------------------|
| LASSO          | 0.768 (0.096)  | 0.630 (0.213)  | 0.366 (0.176)  | 3923 (18)        |
| Adaptive LASSO | 0.787 (0.093)  | 0.622 (0.218)  | 0.381 (0.183)  | 1233 (8)         |
| SCAD           | 0.794 (0.098)  | 0.634 (0.220)  | 0.402 (0.196)  | 674 (12)         |

**4. Monte Carlo simulation.** In this section we use simulations to examine the performance of the penalized log-likelihood approach proposed in Section 2, to estimate the precision matrix with different penalties. In the first three examples, we set the dimensionality $p = 30$. Three different data generating settings for the $30 \times 30$ precision matrix $\boldsymbol{\Omega}$ are considered in Examples 4.1, 4.2 and 4.3. In Examples 4.4 and 4.5 we consider the corresponding high dimensional case with $p = 200$ for Examples 4.1 and 4.2, respectively. In each example we first generate a true precision matrix $\boldsymbol{\Omega}$ which will be fixed for the whole example. Next we generate a dataset of $n = 120$ i.i.d. random vectors distributed as $N(\boldsymbol{0}, \boldsymbol{\Omega}^{-1})$. For each simulated dataset and each penalty a 6-fold cross-validation scheme is used to tune the regularization parameter as discussed in Section 2.3.

To compare the performance of different estimators corresponding to the three penalty functions under consideration, the LASSO, adaptive LASSO and SCAD, we use two types of loss functions: the entropy loss and the quadratic loss [Lin and Perlman (1985)] defined by

$$\text{loss}_1(\boldsymbol{\Omega}, \hat{\boldsymbol{\Omega}}) = \text{tr}\, \boldsymbol{\Omega}^{-1}\hat{\boldsymbol{\Omega}} - \log|\boldsymbol{\Omega}^{-1}\hat{\boldsymbol{\Omega}}| - n \quad \text{and} \quad \text{loss}_2(\boldsymbol{\Omega}, \hat{\boldsymbol{\Omega}}) = \text{tr}(\boldsymbol{\Omega}^{-1}\hat{\boldsymbol{\Omega}} - I)^2,$$

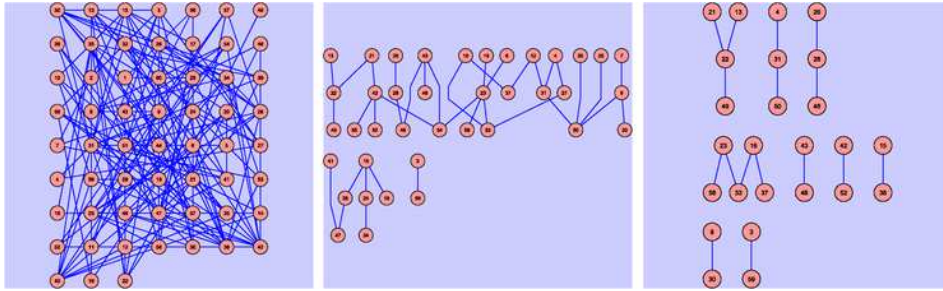

FIG. 2. *Gene networks derived using three penalties: the LASSO (left panel), the adaptive LASSO (middle panel) and the SCAD (right panel).*

TABLE 3
*Simulation result of Example 4.1*

|  | loss$_1$ | loss$_2$ | zero$_1$ | zero$_2$ | perc$_1$ | perc$_2$ |
|---|---|---|---|---|---|---|
| LASSO | 1.64 (0.15) | 11.06(6.64) | 248.48 (60.02) | 0.02 (0.20) | 30.60 (7.39) | 0.02 (0.23) |
| Adaptive LASSO | 1.14 (0.16) | 7.44(4.45) | 42.58 (28.71) | 0.16 (0.56) | 5.24 (3.54) | 0.18 (0.62) |
| SCAD | 0.83 (0.24) | 2.49(3.78) | 76.89 (23.58) | 0.18 (0.58) | 9.47 (2.90) | 0.20 (0.65) |

respectively, where $\hat{\boldsymbol{\Omega}}$ is an estimate of the true precision matrix $\boldsymbol{\Omega}$. To evaluate the performance of the three different penalties concerning sparsity, we report two types of errors regarding zero elements: zero$_1$ means the number of type-I errors (i.e., the true entry of the precision matrix is nonzero but the corresponding estimate is zero) and zero$_2$ the number of type-II errors (i.e., the true entry is zero but its estimator is nonzero). Ideally, we would like to have small zero$_1$ and zero$_2$. We also calculate the relative error percentages: perc$_1$ = $100 \times$ zero$_1$/N$_1$ and perc$_2$ = $100 \times$ zero$_2$/N$_2$, where N$_1$ and N$_2$ are the number of zeros and nonzeros of the true precision matrix respectively. Results of loss$_1$, loss$_2$, zero$_1$, zero$_2$, perc$_1$ and perc$_2$ over the 100 simulations are reported for each simulation example. We will summarize the performance at the end of this section.

EXAMPLE 4.1 [Tridiagonal case ($n = 120$, $p = 30$)].   In this first example we consider the case with a tridiagonal precision matrix, which is associated with the autoregressive process of order one [i.e., AR(1) covariance structure]. In this case the covariance matrix $\boldsymbol{\Sigma}$ is a $p \times p$ matrix with $(i, j)$-element $\sigma_{ij} = \exp(-a|s_i - s_j|)$, where $s_1 < s_2 < \cdots < s_p$ for some $a > 0$. Here, we choose

$$s_i - s_{i-1} \overset{\text{i.i.d.}}{\sim} \text{Unif}(0.5, 1), \qquad i = 2, \ldots, p.$$

The precision matrix is set as $\boldsymbol{\Omega} = \boldsymbol{\Sigma}^{-1}$. The performance of three penalties over 100 repetitions is reported in Table 3, which presents the means of zero$_1$, zero$_2$, loss$_1$, loss$_2$, perc$_1$ and perc$_2$ with their corresponding standard errors in parentheses.

It is not realistic to plot the individual sparsity pattern of the estimates for all the repetitions. Instead we plot the average sparsity pattern, the relative frequency matrix, for each penalty. More specifically, the $(i, j)$-element of the relative frequency matrix is defined as the relative frequency of nonzero estimates of the $(i, j)$-element of the precision matrix $\boldsymbol{\Omega}$ throughout the 100 repetitions. For example, the diagonal elements $\omega_{ii}$ have estimates that are always nonzero and, as a result, their corresponding relative frequencies are always one. We plot this average sparsity pattern using different penalties in panels B, C and D of Figure 3. The true precision matrix is given in panel A of Figure 3. We render this kind of sparsity pattern graph using the gray-scale version of "imagesc" function in Matlab.

EXAMPLE 4.2 [General case ($n = 120$, $p = 30$)]. In the second example we consider a general sparse precision matrix and use the data generating scheme of Li and Gui (2006). More specifically, we generate $p$ points randomly on the unit square and calculate all their pairwise distances. For each point, define its $k$ nearest neighbors as those with $k$ smallest distances to this point. By choosing different number $k$, we can obtain graphs for this model with different degrees of sparsity. For each "edge," the corresponding element in the precision matrix is generated uniformly over $[-1, -0.5] \cup [0.5, 1]$. The value of the $i$th diagonal entry is set as a multiple of the sum of the absolute values of the $i$th row elements excluding the diagonal entry. Here we chose a multiple of 2 to ensure that the obtained precision matrix is positive definite. Finally, each row is divided by the corresponding diagonal element so that the final precision matrix has diagonal elements of ones. Numerical results are summarized in Figure 4 and Table 4.

EXAMPLE 4.3 [Exponential decay matrix ($n = 120$, $p = 30$)]. In this example we consider the case that no element of the precision matrix is exactly zero. The $(i, j)$-element of the true precision matrix is given by $\omega_{ij} = \exp(-2|i - j|)$, which can be extremely small when $|i - j|$ is large. Numerical results over 100 repetitions in the same format as Example 4.1 are reported in Table 5 and Figure 5. Notice in Figure 5, panel A shows the sparsity pattern, since we apply the threshold to the true precision matrix as to the three estimates.

EXAMPLE 4.4 [High dimensional tridiagonal case ($n = 120$, $p = 200$)]. The previous three examples belong to the classical setting with dimensionality $p$ smaller than the sample size $n$. Next we investigate the high dimensional case with $p > n$. In this example we keep all the data generation process of Example 4.1 except that we increase the dimensionality $p$ to 200. The simulation result is reported in Table 6 and Figure 6.

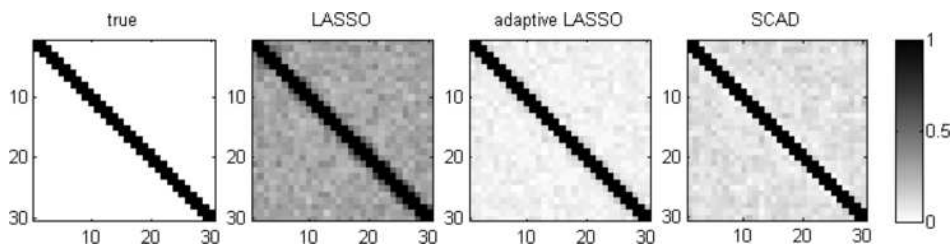

FIG. 3. *For the 100 samples in Example 4.1, the average sparsity pattern recovery for the LASSO, adaptive LASSO and SCAD penalties are plotted in panels* B, C *and* D, *respectively, to compare with the true sparsity pattern (panel* A*).*

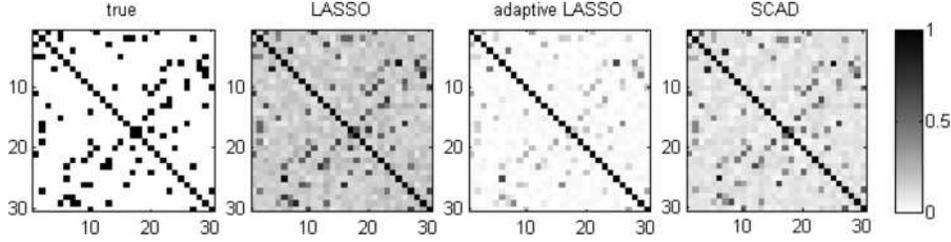

Fig. 4. *For the 100 samples in Example 4.2, the average sparsity pattern recovery for the LASSO, adaptive LASSO and SCAD penalties are plotted in panels* B, C *and* D, *respectively, to compare with the true sparsity pattern (panel* A*).*

EXAMPLE 4.5 [High dimensional general case ($n = 120$, $p = 200$)]. In this example we use the same setting as that of Example 4.2 but increase $p$ to 200 as we did in Example 4.4. The simulation results are summarized in Table 7 and Figure 7.

Throughout all these different examples, we can see that the LASSO penalty, in general, produces more nonzero elements in the estimated precision matrix than the adaptive LASSO and SCAD penalties. This is due to the bias inherited in the LASSO penalty that prevents data from choosing a large regularization parameter. The adaptive LASSO produces the most sparse pattern due to the specific choice of the initial estimate. Based on

TABLE 4
*Simulation result of Example 4.2*

|  | loss$_1$ | loss$_2$ | zero$_1$ | zero$_2$ | perc$_1$ | perc$_2$ |
|---|---|---|---|---|---|---|
| LASSO | 1.11 (0.11) | 9.05 (4.35) | 125.66 (39.79) | 34.62 (8.28) | 15.99 (5.06) | 30.37 (7.26) |
| Adaptive LASSO | 1.14 (0.10) | 2.99 (2.17) | 11.28 (10.35) | 66.80 (8.53) | 1.44 (1.32) | 58.60 (7.48) |
| SCAD | 1.04 (0.10) | 0.81 (1.12) | 62.72 (26.79) | 45.96 (9.35) | 7.98 (3.41) | 40.32 (8.20) |

TABLE 5
*Simulation result of Example 4.3*

|  | loss$_1$ | loss$_2$ | zero$_1$ | zero$_2$ | perc$_1$ | perc$_2$ |
|---|---|---|---|---|---|---|
| LASSO | 0.88 (0.09) | 10.72 (4.93) | 88.54 (34.33) | 126.94 (12.57) | 12.61 (4.89) | 64.11 (6.35) |
| Adaptive LASSO | 0.81 (0.07) | 4.25 (2.93) | 5.08 (6.71) | 161.62 (6.16) | 0.72 (0.96) | 81.63 (3.11) |
| SCAD | 0.75 (0.08) | 0.77 (1.07) | 35.60 (23.03) | 145.28 (12.09) | 5.07 (3.28) | 73.37 (6.11) |

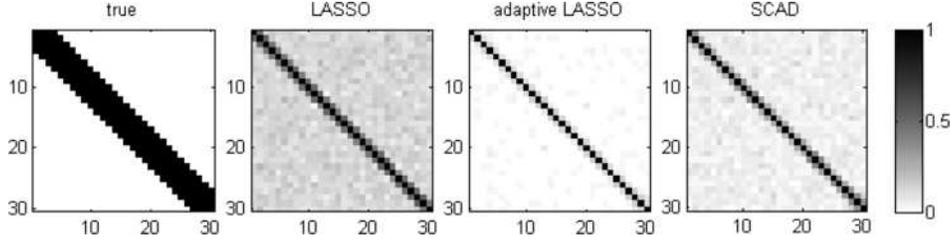

Fig. 5. *For the 100 samples in Example 4.3, the average sparsity pattern recovery for the LASSO, adaptive LASSO and SCAD penalties are plotted in panels* B, C *and* D, *respectively, to compare with the true sparsity pattern (panel* A*).*

TABLE 6
*Simulation result of Example 4.4*

|          | loss$_1$ | loss$_2$ | zero$_1$ | zero$_2$ | perc$_1$ | perc$_2$ |
|----------|----------|----------|----------|----------|----------|----------|
| LASSO    | 19.31 (0.43) | 1065.37 (82.56) | 4009.75 (117.60) | 0.64 (1.24) | 10.18 (0.30) | 0.11 (0.21) |
| Adaptive LASSO | 12.44 (0.92) | 664.46 (129.35) | 269.86 (61.97) | 7.76 (4.11) | 0.68 (0.16) | 1.30 (0.69) |
| SCAD     | 10.55 (0.48) | 288.26 (62.34) | 3478.76 (106.73) | 1.10 (1.67) | 8.83 (0.27) | 0.18 (0.28) |

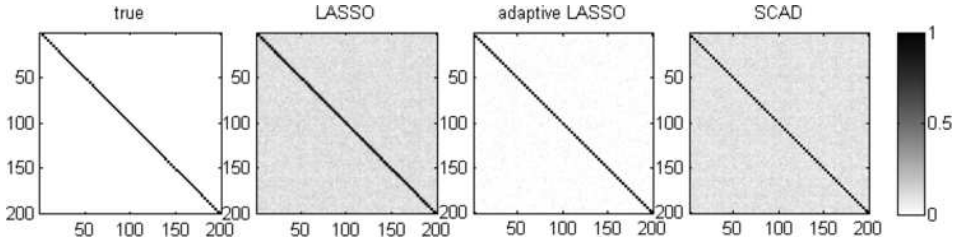

Fig. 6. *For the 100 samples in Example 4.4, the average sparsity pattern recovery for the LASSO, adaptive LASSO and SCAD penalties are plotted in panels* B, C *and* D, *respectively, to compare with the true sparsity pattern (panel* A*).*

Tables 3–7, improvements are observed for the adaptive LASSO and SCAD penalties over the LASSO penalty in terms of the two types of loss functions (especially the second type) and as well as the two types of errors regarding zero elements.

## 5. Theoretical properties.
In this section we provide some theoretical justifications. We first prove that the penalized log-likelihood function is increasing in each iteration using the LLA algorithm. The oracle properties of the SCAD and adaptive LASSO penalties will be established in our context.

Without loss of generality, we may consider the case that the random vector is normally distributed with mean zero, that is, $\mathbf{X} \sim N(\mathbf{0}, \mathbf{\Sigma}_0)$, where

TABLE 7
*Simulation result of Example 4.5*

|  | loss$_1$ | loss$_2$ | zero$_1$ | zero$_2$ | perc$_1$ | perc$_2$ |
|---|---|---|---|---|---|---|
| LASSO | 8.24 (0.27) | 1082.61 (112.61) | 796.16 (264.66) | 255.22 (13.57) | 2.02 (0.67) | 46.74 (2.49) |
| Adaptive LASSO | 6.50 (0.21) | 316.95 (53.99) | 6.58 (4.92) | 336.24 (4.51) | 0.02 (0.01) | 61.58 (0.83) |
| SCAD | 6.65 (0.40) | 32.33 (23.06) | 224.98 (247.45) | 298.12 (21.24) | 0.57 (0.63) | 54.60 (3.89) |

**0** is a vector of zeros and $\mathbf{\Sigma}_0$ is the true unknown $p \times p$ covariance matrix. The corresponding true precision matrix is $\mathbf{\Omega}_0 = \mathbf{\Sigma}_0^{-1}$. Our sample consists of $n$ independent and identically distributed observations $\mathbf{x}_1, \mathbf{x}_2, \ldots, \mathbf{x}_n$. In this case the sample covariance matrix is defined by

$$(5.1) \qquad \hat{\mathbf{\Sigma}} = \sum_{i=1}^{n} \mathbf{x}_i \mathbf{x}_i^T / n.$$

Note here $p$ is assumed to be fixed and we study asymptotic properties of our penalized estimates with the SCAD and adaptive LASSO penalties as the sample size $n \to \infty$.

THEOREM 5.1. *For a differentiable concave penalty function $p_\lambda(\cdot)$ on $[0, \infty]$, the penalized log-likelihood function is increasing through each iteration in the LLA approximation.*

See the Appendix for the proof of Theorem 5.1.

THEOREM 5.2. *For $n$ i.i.d. observations $\mathbf{x}_1, \mathbf{x}_2, \ldots, \mathbf{x}_n$ from $N(\mathbf{0}, \mathbf{\Sigma}_0)$, the optimizer $\hat{\mathbf{\Omega}}$ of the SCAD penalized log-likelihood function (2.4) with sample covariance given by (5.1) has the oracle property in the sense of Fan and Li (2001), when $\lambda \to 0$ and $\sqrt{n}\lambda \to \infty$ as $n \to \infty$. Namely:*

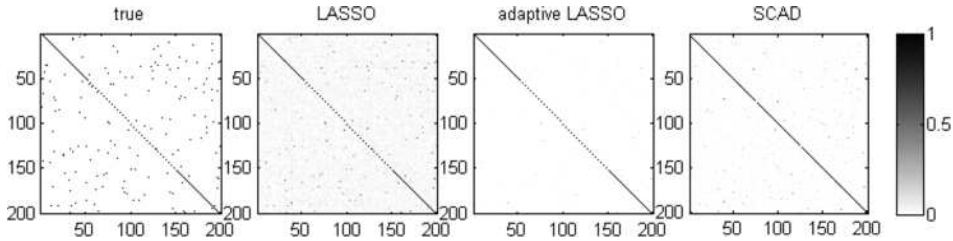

FIG. 7. *For the 100 samples in Example 4.5, the average sparsity pattern recovery for the LASSO, adaptive LASSO and SCAD penalties are plotted in panels B, C and D, respectively, to compare with the true sparsity pattern (panel A).*

(1) *Asymptotically, the estimate $\hat{\boldsymbol{\Omega}}$ has the same sparsity pattern as the true precision matrix $\boldsymbol{\Omega}_0$.*

(2) *The nonzero entries of the $\hat{\boldsymbol{\Omega}}$ are $\sqrt{n}$-consistent and asymptotically normal.*

See the Appendix for the proof of Theorem 5.2.

THEOREM 5.3.   *When $\sqrt{n}\lambda = O_p(1)$ and $\lambda\sqrt{n}a_n^\gamma \to \infty$ as $n \to \infty$, the oracle property also holds for the adaptive LASSO penalty with weights specified by $w_{ij} = 1/|\tilde{w}_{ij}|^\gamma$ for some $\gamma > 0$ and any $a_n$-consistent estimator $\tilde{\boldsymbol{\Omega}} = (\tilde{\omega}_{ij})_{1\le i,j \le p}$, that is, $a_n(\tilde{\boldsymbol{\Omega}} - \boldsymbol{\Omega}_0) = O_p(1)$.*

The proof of Theorem 5.3 can be found in the supplemental article [Fan, Feng and Wu (2008)].

## APPENDIX

PROOF OF THEOREM 5.1.   Define

$$Q_\lambda(\boldsymbol{\Omega}) = \log\det\boldsymbol{\Omega} - \langle\hat{\boldsymbol{\Sigma}}, \boldsymbol{\Omega}\rangle - \sum_{i=1}^{n}\sum_{j=1}^{n}p_\lambda(|\omega_{ij}|)$$

and

$$\Phi_\lambda(\boldsymbol{\Omega}|\hat{\boldsymbol{\Omega}}) = \log\det\boldsymbol{\Omega} - \langle\hat{\boldsymbol{\Sigma}}, \boldsymbol{\Omega}\rangle - \sum_{i=1}^{n}\sum_{j=1}^{n}[p_\lambda(|\hat{\omega}_{ij}|) + p_\lambda'(|\hat{\omega}_{ij}|)(|\omega_{ij}| - |\hat{\omega}_{ij}|)].$$

Then, given estimate $\hat{\boldsymbol{\Omega}}^{(k)}$, we have

$$(A.1) \qquad \hat{\boldsymbol{\Omega}}^{(k+1)} = \arg\max_{\boldsymbol{\Omega}\in S_p}\Phi_\lambda(\boldsymbol{\Omega}|\hat{\boldsymbol{\Omega}}^{(k)}).$$

Our goal is to prove that $Q_\lambda(\hat{\boldsymbol{\Omega}}^{(k+1)}) \ge Q_\lambda(\hat{\boldsymbol{\Omega}}^{(k)})$. At the $k$th-step, consider

$$Q_\lambda(\boldsymbol{\Omega}) - \Phi_\lambda(\boldsymbol{\Omega}|\hat{\boldsymbol{\Omega}}^{(k)})$$
$$= \sum_{i=1}^{n}\sum_{j=1}^{n}\{p_\lambda(|\hat{\omega}_{ij}^{(k)}|) + p_\lambda'(|\hat{\omega}_{ij}^{(k)}|)(|\omega_{ij}| - |\hat{\omega}_{ij}^{(k)}|) - p_\lambda(|\omega_{ij}|)\}.$$

By the concavity of $p_\lambda(\cdot)$ over $[0,\infty)$, we have $p_\lambda(|\hat{\omega}_{ij}^{(k)}|) + p_\lambda'(|\hat{\omega}_{ij}^{(k)}|)(|\omega_{ij}| - |\hat{\omega}_{ij}^{(k)}|) - p_\lambda(|\omega_{ij}|) \ge 0$. Then, we have $Q_\lambda(\boldsymbol{\Omega}) \ge \Phi_\lambda(\boldsymbol{\Omega}|\hat{\boldsymbol{\Omega}}^{(k)})$. Finally, by noticing that $Q_\lambda(\hat{\boldsymbol{\Omega}}^{(k)}) = \Phi_\lambda(\hat{\boldsymbol{\Omega}}^{(k)}|\hat{\boldsymbol{\Omega}}^{(k)})$ and using (A.1), we have

$$Q_\lambda(\hat{\boldsymbol{\Omega}}^{(k+1)}) \ge \Phi_\lambda(\hat{\boldsymbol{\Omega}}^{(k+1)}|\hat{\boldsymbol{\Omega}}^{(k)}) \ge \Phi_\lambda(\hat{\boldsymbol{\Omega}}^{(k)}|\hat{\boldsymbol{\Omega}}^{(k)}) = Q_\lambda(\hat{\boldsymbol{\Omega}}^{(k)}),$$

as desired. $\square$

PROOF OF THEOREM 5.2. It is enough to check conditions (A)–(C) of Fan and Li ([2001]). Since $\mathbf{x}_i$ are i.i.d. from $N(\mathbf{0}, \boldsymbol{\Sigma}_0)$, the probability density function for $\mathbf{X}$ is given by $f(\mathbf{x}, \boldsymbol{\Omega}_0) = \exp(-\mathbf{x}^T \boldsymbol{\Omega}_0 \mathbf{x}/2) \sqrt{\det \boldsymbol{\Omega}_0/(2\pi)}$. The log-likelihood function of the precision matrix is given by

$$\sum_{i=1}^{n} \frac{1}{2} (\log \det \boldsymbol{\Omega} - \mathbf{x}_i^T \boldsymbol{\Omega} \mathbf{x}_i)$$

$$= \frac{n}{2} \left( \log \det \boldsymbol{\Omega} - \frac{1}{n} \sum_{i=1}^{n} \mathbf{x}_i^T \boldsymbol{\Omega} \mathbf{x}_i \right)$$

$$= \frac{n}{2} (\log \det \boldsymbol{\Omega} - \operatorname{tr}(\boldsymbol{\Omega} \hat{\boldsymbol{\Sigma}})),$$

up to a constant, where $\operatorname{tr}(\cdot)$ denotes the trace operator. This justifies the log-likelihood function given in Section 2 as well.

Notice that

$$E_{\boldsymbol{\Omega}_0} \left( \frac{\partial \log f(\mathbf{x}, \boldsymbol{\Omega})}{\partial \omega_{ij}} \right) \Big|_{\boldsymbol{\Omega} = \boldsymbol{\Omega}_0} = \frac{1}{2} E_{\boldsymbol{\Omega}} \frac{\partial}{\partial \omega_{ij}} (\log \det \boldsymbol{\Omega} - \mathbf{x}^T \boldsymbol{\Omega} \mathbf{x}) \mid_{\boldsymbol{\Omega} = \boldsymbol{\Omega}_0},$$

which reduces to $(-1)^{i+j} \det \boldsymbol{\Omega}_{0,-ij}/(\det \boldsymbol{\Omega}_0) - \sigma_{0,ij}$ when $i \neq j$ and $\frac{1}{2} (\det \boldsymbol{\Omega}_{0,-ii}/(\det \boldsymbol{\Omega}_0) - \sigma_{0,ii})$ when $i = j$, where $\boldsymbol{\Omega}_{0,-ij}$ denotes the matrix after removing the $i$th row and $j$th column from $\boldsymbol{\Omega}_0$ and $\sigma_{0,ij}$ is the $(i, j)$-element of the covariance matrix $\boldsymbol{\Sigma}_0$. Noting that $\boldsymbol{\Omega}_0 = \boldsymbol{\Sigma}_0^{-1}$, we have $(-1)^{i+j} \det \boldsymbol{\Omega}_{0,-ij}/(\det \boldsymbol{\Omega}_0) - \sigma_{0,ij} = 0$ for $i \neq j$ and $\frac{1}{2} (\det \boldsymbol{\Omega}_{0,-ii}/(\det \boldsymbol{\Omega}_0) - \sigma_{0,ii}) = 0$ when $i = j$, as we have desired. That is, $E_{\boldsymbol{\Omega}_0} (\frac{\partial \log f(\mathbf{x}, \boldsymbol{\Omega})}{\partial \omega_{ij}}) |_{\boldsymbol{\Omega} = \boldsymbol{\Omega}_0} = 0$. Similarly, we can show that $E_{\boldsymbol{\Omega}_0} (\frac{\partial}{\partial \omega_{ij}} \log f(\mathbf{x}, \boldsymbol{\Omega}) \frac{\partial}{\partial \omega_{kl}} \log f(\mathbf{x}, \boldsymbol{\Omega})) |_{\boldsymbol{\Omega} = \boldsymbol{\Omega}_0} = E_{\boldsymbol{\Omega}_0} (-\frac{\partial^2}{\partial \omega_{ij} \partial \omega_{kl}} \log f(\mathbf{x}, \boldsymbol{\Omega})) |_{\boldsymbol{\Omega} = \boldsymbol{\Omega}_0}$. So condition (A) is satisfied by noting that $f(\mathbf{x}, \boldsymbol{\Omega})$ has a common support and the model is identifiable.

To prove condition (B), it is sufficient to prove that the log-det function is concave. More explicitly, for the log-det function $h(\boldsymbol{\Omega}) = \log \det \boldsymbol{\Omega}$, we can verify concavity by considering an arbitrary line, given by $\boldsymbol{\Omega} = Z + tV$, where $Z, V \in S_p$. We define $g(t) = h(Z + tV)$, and restrict $g$ to the interval of values of $t$ for which $Z + tV \in S_p$. Without loss of generality, we can assume $t = 0$ is inside the interval, that is, $Z \in S_p$. We have

$$g(t) = \log \det(Z + tV)$$

$$= \log \det(Z^{1/2}(I + tZ^{-1/2} V Z^{-1/2}) Z^{1/2})$$

$$= \sum_{i=1}^{p} \log(1 + t\lambda_i) + \log \det Z,$$

where $\lambda_1, \ldots, \lambda_p$ are the eigenvalues of $Z^{-1/2}VZ^{-1/2}$. Therefore, we have

$$g'(t) = \sum_{i=1}^{p} \frac{\lambda_i}{1+t\lambda_i}, g''(t) = -\sum_{i=1}^{p} \frac{\lambda_i^2}{(1+t\lambda_i)^2}.$$

Since $g''(t) \le 0$, we conclude that $h$ is concave.

Condition (C) is easy to satisfy because the third order derivative does not involve $\mathbf{x}$. □

**Acknowledgments.** The authors thank the Editor, the Associate Editor and two referees, whose comments have greatly improved the scope and presentation of the paper. The authors are in deep debt to Professor d'Aspremont for his helpful discussion and Dr. Alexander McLain for careful proofreading.

## SUPPLEMENTARY MATERIAL

**Proof of Theorem 5.3** (DOI: 10.1214/08-AOAS215SUPP; .pdf). We gave a detailed proof of the oracle properties for the adaptive lasso penalty as stated in Theorem 5.3.

J. Fan
Y. Feng
Department of Operations Research
    and Financial Engineering
Princeton University
Princeton, New Jersey 08544
USA

Y. Wu
Department of Statistics
North Carolina State University
Raleigh, North Carolina 27695
USA